\documentclass[10pt,aps,prl,twocolumn,superscriptaddress, nofootinbib]{revtex4-1}
\usepackage{amsmath,amssymb,mathdots, graphicx}
\newcommand{\ket}[1]{\rvert#1\rangle}
\newcommand{\bra}[1]{\langle #1\rvert}

\begin{document}

\title{A novel analytic spin chain model with fractional revival}
\author{Jean-Michel Lemay}
\email[]{jean-michel.lemay.1@umontreal.ca}
\author{Luc Vinet}
\email[]{luc.vinet@umontreal.ca}
\author{Alexei Zhedanov}
\email[]{zhedanov@yahoo.com}
\thanks{On leave of absence from Donetsk Institute for Physics and Technology, Donetsk 83114, Ukraine}
\affiliation{Centre de recherches math\'ematiques, Universit\'e de Montr\'eal, P.O. Box 6128, Centre-ville Station, Montr\'eal (Qu\'ebec) H3C 3J7}
\date{\today}

\begin{abstract}
  New analytic spin chains with fractional revival are introduced. Their nearest-neighbor couplings and local magnetic fields correspond to the recurrence coefficients of para-Racah polynomials which are orthogonal on quadratic bi-lattices. These models generalize the spin chain associated to the dual-Hahn polynomials. Instances where perfect state transfer also occurs are identified.
\end{abstract}
\maketitle

\section{1. Introduction}
  Fractional revival (FR) is observed when clones of a wave packet reproduce with periodicities in a localized fashion \cite{2004_Robinett_PhysReports_392_1}. This phenomenon has been shown to occur in XX spin chains with pre-engineered couplings. Specifically, in such instances, a state with a spin up initially at one end of the chain evolves after some time $T$ into a state for which the amplitude to find the spin up at a given site is non zero only for the two ends of the chain \cite{2007_Chen&Song&Sun_PhysRevA_75_012113, 2010_Dai_Feng_Kwek_JPhysA, 2010_Kay_IntJQtmInf_8_641, 2015_Banchi&Compagno&Bose_PhysRevA_91_052323, 2015_Exact_Fractional_Revival, Fractional_quantum_spin_chains}. The special case when the spin up localizes exclusively at the end of the chain at some time $T$ is referred to as perfect state transfer (PST). (See \cite{2012_Vinet&Zhedanov_PhysRevA_85_012323} and references therein.) This has cast XX spin chains as attractive systems for the design of wires that can transport quantum information with high fidelity, generate entanglement or provide remote gates. The interest in these questions is generating an abundant literature (See \cite{2007_Bose_ContempPhys_48_13, 2010_Kay_IntJQtmInf_8_641} for reviews and further references). A key advantage of these models is that the tasks are performed through the chain dynamics without the need for external control operations beyond the input/output interventions.

  A systematic analysis of FR at two sites in XX spin chains has been carried in \cite{Fractional_quantum_spin_chains}. In general, FR is essentially described by two parameters that can be tuned independently. The first is connected to a prescribed one-parameter deformation \cite{2015_Exact_Fractional_Revival}. The second comes from a remarkable analytic model. Let us briefly explains its features.

  For the purpose of studying fractional revival it suffices to consider states with only one spin up. When restricted to the one-excitation sector, the XX spin chain Hamiltonians with nearest-neighbor couplings become tridiagonal matrices $J$ that are diagonalized by polynomials orthogonal on the finite set of points formed by the eigenvalues of $J$. One necessary condition for PST is that the matrices $J$ possess a special property called mirror-symmetry \cite{2004_Albanese&Christandl&Datta&Ekert_PhysRevLett_93_230502}. If one looks for fractional revival in systems with special Hamiltonians whose restrictions $J$ are mirror-symmetric, one generically obtains analytic models where the couplings and magnetic fields are exactly given by the recurrence coefficients of polynomials that are orthogonal on linear bi-lattices. By a linear bi-lattice we mean the set of points obtained by shifting two equally-spaced linear lattices with respect to one another: 
  \begin{align} \label{1}
  x_s = x_0 + s +\frac{1}{2}(\delta-1)(1-(-1)^s)\quad s=0,\dots,N. 
  \end{align}
  We assume that the chains have $N+1$ sites. Interestingly, the associated orthogonal polynomials have only been discovered recently in the context of PST studies \cite{2012_Vinet&Zhedanov_JPhysA_45_265304}. They have been called para-Krawtchouk polynomials. The corresponding FR parameter is related to the relative shift $\delta$.

  Now there is a procedure known as spectral surgery \cite{2012_Vinet&Zhedanov_PhysRevA_85_012323} that allows to remove spectral points while preserving mirror-symmetry. In principle this permits to obtain any prescribed spectrum for $J$ and the corresponding mirror-symmetric couplings and magnetic fields by removing in this way the appropriate elements from the linear bi-lattice set. As explained in \cite{Fractional_quantum_spin_chains}, upon performing the isospectral deformation (mentioned above) of surgered models, one may construct all XX spin chains with FR. 

  The usefulness of exactly solvable models does not need to be stressed. They make possible the analytic exploration of the dynamics and are rooted in a secular tradition in theoretical physics. Now it is a fact that the repeated removal of energy levels will yield expressions that are more and more complicated for the chain data and will thus have the effect of obscuring the analytic properties of the system. There is hence much interest in finding directly other manifestly analytic models with FR and this is the purpose of this paper.

  The discovery of the para-Krawtchouk polynomials orthogonal on linear bi-lattices has prompted the search for polynomials orthogonal on quadratic bi-lattices. These functions have been found very recently \cite{Para_Racah_Polynomials}. They have been called the para-Racah polynomials and are denoted by $P_n(y^2;N; a, c, \alpha)$. As the notation suggests, they depend on 3 parameters, two of which being related to the definition of the grid or bi-lattice, in addition to the natural number $N$.  

  We shall here discuss the fractional revival properties of the XX spin chains associated to these orthogonal polynomials. Mirror-symmetry requires $\alpha = \frac{1}{2}$. When this is so, FR will occur provided $a$ and $c$ are expressed in terms of solutions of a quadratic Diophantine equation. PST will happen for a subset of the values of the parameters for which FR is realized. When $c = a +\frac{1}{2}$, the models will be seen to reduce to the ones associated to the dual-Hahn polynomials - a paradigm example of chains exhibiting PST \cite{2004_Albanese&Christandl&Datta&Ekert_PhysRevLett_93_230502}. When an isospectral transformation is applied, mirror-symmetry is broken and more general models with FR are obtained and found to correspond to para-Racah polynomials now with an arbitrary $\alpha$.

  The paper will unfold as follows. In Section 2, relevant aspects of fractional revival in XX spin chains will be reviewed. The models with mirror-symmetric couplings and magnetic fields corresponding to the recurrence coefficients of the para-Racah polynomials with $\alpha = \frac{1}{2}$ will be  introduced in Section 3 for $N$ odd, that is for an even number of sites. The conditions (on the parameters) for fractional revival to take place will be determined and the situations of PST will also be identified. The presentation of the models for $N$ even will be carried out via the spectral surgery procedure. This last point will be the object of Section 3 where we shall give the couplings and magnetic fields that result when the last level of the chains with $N$ odd is removed. That the models reduce to the one associated with the dual-Hahn polynomials will be discussed in Section 5. The shape of the couplings and magnetic fields will be depicted in plots exhibiting the differences between the odd and even $N$ cases as well as with the dual-Hahn polynomials situation. The isospectral deformation of the mirror-symmetric chains considered up to that point will be carried out in Section 6 to find analytic models with FR corresponding to the general para-Racah polynomials. The classification will be done in full generality for spin chains with $N>4$. We shall sum up to conclude. 

\section{2. Fractional Revival and Orthogonal Polynomials}
  We shall consider $XX$ spin chains with $N+1$ sites and nearest-neighbor interactions governed by Hamiltonians $H$ of the form
  \begin{equation} \label{Hamiltonian}
  H=\frac{1}{2}\sum_{n=0}^{N-1}J_{n+1}(\sigma_{n}^{x}\sigma_{n+1}^{x}+\sigma_{n}^{y}\sigma_{n+1}^{y})+\frac{1}{2}\sum_{n=0}^{N}B_{n}(\sigma_{n}^{z}+1).
  \end{equation}
  $J_{n}$ is the coupling constant between the sites $n-1$ and $n$ and $B_{n}$ is the magnetic field strength at the site $n$, with $n=0,1,\ldots, N$. The symbols $\sigma_{n}^{x}$, $\sigma_{n}^{y}$, $\sigma_{n}^{z}$ stand for the Pauli matrices with the index $n$ indicating on which $\mathbb{C}^2$ copy of $(\mathbb{C}^2)^{\otimes N+1}$ they act. It is easy to see that the Hamiltonian $H$ is invariant under rotation around the z-axis:  
  \begin{align}
  \left[H,\frac{1}{2}\sum_{n=0}^{N}(\sigma_{n}^{z}+1)\right]=0.
  \end{align}
  This implies that the eigenstates of $H$ split in subspaces labeled by the number of spins over the chain that are up, i.e. that are eigenstates of $\sigma^{z}$ with eigenvalue $+1$. To study fractional revival it suffices to focus on the restriction $J$ of $H$ to the one-excitation sector. The states of that subspace are naturally described by the canonical basis vectors of $\mathbb{Z}^{\otimes N+1}$
  \begin{align}
  \ket{n}=(0,0,\ldots, 1,\ldots, 0)^\intercal
  \end{align}
  with the single $1$ in the $n^{th}$ position corresponding to the single spin up at the $n^{th}$ site. The action of $J$ in that basis follows from \eqref{Hamiltonian} and is given by
  \begin{align} \label{Jrec}
  J\ket{n}=J_{n+1}\ket{n+1}+B_{n}\ket{n}+J_{n}\ket{n-1}
  \end{align}
  where $J_0=J_{N+1}=0$ is assumed. That is, $J$ is the following Jacobi matrix
  \begin{align} \label{JacobiMatrix}
  J=
  \begin{pmatrix}
  B_0	&	J_1	&		&	&
  \\
  J_1	&	B_1	&	J_2	&	&
  \\
	  &	J_2	&	B_2	& \ddots &
  \\
	  & 		& \ddots		& 	\ddots & J_{N}
  \\
  & & &J_{N}&B_{N}
  \end{pmatrix}.
  \end{align}
  It will be said to be mirror-symmetric with respect to the anti-diagonal or persymmetric if
  \begin{align} \label{persymmetry2}
  RJR=J
  \end{align}
  with
  \begin{align}
  R = 
  \begin{pmatrix}
    & & & 1  \\
    & & 1&   \\
    & \iddots & &  \\
    1& & &  \\
  \end{pmatrix}.
  \end{align}
  In terms of the couplings and magnetic field strengths, this amounts to
  \begin{align} \label{persymmetry}
  J_n = J_{N+1-n}, \quad B_n=B_{N-n}.
  \end{align}
  Since $B_n, J_n\in\mathbb{R}$, $J$ is clearly hermitian and has real eigenvalues. We can introduce the eigenbasis of $J$ :
  \begin{align}
  J \ket{x_s} = x_s \ket{x_s},
  \end{align}
  where the eigenvalues $x_s$ are assumed to be non-degenerate and are taken to be in increasing order $x_0<x_1<\dots<x_N$. Then, in view of \eqref{Jrec}, it is easy to show that we have the following expansions
  \begin{align} \label{poly_expansions}
  \ket{x_s}=\sum_{n=0}^{N}\sqrt{w_s}\chi_n(x_s)\ket{n} \\
  \ket{n}=\sum_{s=0}^{N}\sqrt{w_s}\chi_n(x_s)\ket{x_s}
  \end{align}
  where $\chi_n(x)$ are orthonormal polynomials on the finite set of spectral points $x_s$ that obey the recurrence relation 
  \begin{align} \label{poly_relrec}
    x\chi_n(x)=J_{n+1}\chi_{n+1}(x)+B_{n}\chi_{n}(x)+J_{n}\chi_{n-1}(x).
  \end{align}
  It can be shown \cite{2012_Vinet&Zhedanov_PhysRevA_85_012323} that $J$ is persymmetric if and only if
  \begin{align} \label{14}
  \chi_N(x_s) = (-1)^{N+s}, \quad s=0,1,\dots,N.
  \end{align}
  Let us now come to fractional revival at two sites. A wave packet initially localized at the site $0$ will be revived at sites $0$ and $N$ after time $T$ if
  \begin{align} \label{15}
  e^{-iTJ}\ket{0} = \xi\ket{0} + \eta\ket{N}
  \end{align}
  with $|\xi|^2+|\eta|^2 = 1$. Given the normalization condition and the inconsequential overall phase, we see that fractional revival at two sites is essentially characterized by two real angles. Observe that we are in a situation of perfect state transfer when $\xi=0$ :
  \begin{align*}
  e^{-iTJ}\ket{0} = e^{i\phi'}\ket{N}.
  \end{align*}
  With the help of expansion \eqref{poly_expansions}, it is immediate to see that condition \eqref{15} translates into
  \begin{align} \label{16}
  e^{-iTx_s} = \xi + \eta \chi_N(x_s).
  \end{align}
  Note that for the right hand side of \eqref{16} to have modulus 1, we must have
  \begin{align} \label{17}
  \chi_N^2(x_s) + 2 \frac{\text{Re}(\xi\eta^*)}{|\eta|^2} = 1.
  \end{align}
  The implications of \eqref{16} have been examined in \cite{Fractional_quantum_spin_chains} to obtain a characterization of the chains with FR. The analysis proceeds in two steps. Condition \eqref{16} is first enforced when $\xi$ and $\eta$ involve only one of the two essential angles (apart from the global phase $\phi$) and are expressed like this : 
  \begin{align} \label{18}
   \xi = e^{i\phi}\sin2\theta \quad \eta = ie^{i\phi}\cos2\theta.
  \end{align}
  The additional parameter is then introduced by subsequently performing an isospectral deformation of the Jacobi matrix obtained as a result of the first step. This course will be followed here.

  When $\xi$ and $\eta$ take the special form \eqref{18}, we see from \eqref{17} that $\chi_N^2(x_s) = 1$. Simple considerations (explained fully in \cite{Fractional_quantum_spin_chains}) lead one to conclude that condition \eqref{14} must then be obeyed. In other words, the fractional revival condition \eqref{16} requires that $J$ be persymmetric when $\xi$ and $\eta$ are as in \eqref{18}. Moreover, with these expressions for $\xi$ and $\eta$, the real and imaginary parts of \eqref{16} yield the following relations that the eigenvalues $x_s$ of $J$ must satisfy :
  \begin{subequations} \label{19}
  \begin{align}
  &\cos(Tx_s+\phi) = \cos\left(\frac{\pi}{2}-2\theta\right) \\
  &\sin(Tx_s+\phi) = (-1)^{N+s+1}\sin\left(\frac{\pi}{2}-2\theta\right).
  \end{align}
  \end{subequations}
  The determination of $J$ and of $H$ as a result, is henceforth framed as an inverse spectral problem that can be solved using the theory of orthogonal polynomials. The eigenvalues $x_s$ found to verify the FR conditions \eqref{19} determine the characteristic polynomial of degree $N+1$ and the knowledge of $\chi_N$ at the $N+1$ points $x_s$, as prescribed by \eqref{14}, completely specifies this polynomial also. All the other polynomials $\chi_n(x)$ can be constructed from these two by using the Euclidian algorithm (see \cite{2012_Vinet&Zhedanov_PhysRevA_85_012323}) and this gives $J$ and $H$.

  The generic set of eigenvalues satisfying conditions \eqref{19} is a linear bi-lattice and the algorithm we just explained leads in this case to the para-Krawtchouk polynomials \cite{2012_Vinet&Zhedanov_JPhysA_45_265304}. The specifications of the chain are then provided by the recurrence coefficients. 

  In the following, we shall not adopt this deductive approach which is explained fully in \cite{Fractional_quantum_spin_chains}. We shall rather identify a new analytic model by proceeding in the reverse. We shall first provide a set of mirror-symmetric couplings and magnetic fields known to form the recurrence coefficients of polynomials that are orthogonal with respect to quadratic bi-lattices. We shall then determine for what values of the grid parameters are the FR conditions \eqref{19} satisfied. The isospectral deformations that allow to relax the condition that $J$ is persymmetric will be presented in the last section.  

\section{3. A MODEL WITH FRACTIONAL REVIVAL BASED ON PARA-RACAH POLYNOMIALS FOR N ODD}
  The analytic model introduced here is based on the para-Racah polynomials $P_n(y^2;N;a,c,\alpha)$. These polynomials have only been identified recently \cite{Para_Racah_Polynomials}. For now, take $N$ to be odd and write
  \begin{align} \label{20}
  N=2j+1.
  \end{align}
  When $\alpha=\frac{1}{2}$, the recurrence coefficients of the para-Racah polynomials provide the following explicit expressions for couplings and magnetic fields :
  \begin{align} \label{Couplings}
    B_n&=\frac{1}{2}\left[a(a+j)+c(c+j)+n(N-n)\right], \\ \notag
    J_n&=\bigg[\frac{n(N+1-n)(N-n+a+c)(n-1+a+c)}{4(N-2n)(N-2n+2)}\\ &\times\left((n-j-1)^2-(a-c)^2\right)\bigg]^{\frac{1}{2}},\notag
  \end{align}
  where the parameters $a$ and $c$ are such that $a>-\frac{1}{2}$ and $|a|<c<|a+1|$. It can be checked directly that these $J_n$ and $B_n$ satisfy \eqref{persymmetry} and that they hence form a persymmetric matrix.

  The para-Racah polynomials are known \cite{Para_Racah_Polynomials} to satisfy a discrete orthogonality relation on the points of the quadratic bi-lattice defined by 
  \begin{align} \label{bi-lattice}
  \begin{aligned}
    x_{2s}=(s+a)^2, \quad s=0,\dots,j, \\
    x_{2s+1}=(s+c)^2, \quad s=0,\dots,j.
  \end{aligned}
  \end{align}    
  Now, FR will be observed when such eigenvalues obey conditions \eqref{19}. Splitting the even and odd cases and using $N=2j+1$, FR requires that
  \begin{subequations}
  \begin{align}
  \cos(T(s+a)^2+\phi)&=\cos\left(\frac{\pi}{2}-2\theta\right),\\
  \sin(T(s+a)^2+\phi)&=\sin\left(\frac{\pi}{2}-2\theta\right),\\
  \cos(T(s+c)^2+\phi)&=\cos\left(-\frac{\pi}{2}+2\theta\right),\\
  \sin(T(s+c)^2+\phi)&=\sin\left(-\frac{\pi}{2}+2\theta\right), 
  \end{align}
  \end{subequations} 
  for $s=0,1,\dots,j$. This implies that
  \begin{subequations} \label{arguments}
  \begin{align}
  T(s+a)^2&=\frac{\pi}{2}-2\theta-\phi+2\pi M_s, \\
  T(s+c)^2&=-\frac{\pi}{2}+2\theta-\phi+2\pi L_s, 
  \end{align}
  \end{subequations} 
  where $M_s$ and $L_s$ are arbitrary sequences of integers. Now, since the LHS of \eqref{arguments} are quadratic functions of $s$, the RHS must also be quadratic functions of $s$. Hence, $M_s$ and $L_s$ take the general form 
  \begin{subequations} \label{int_sequences}
  \begin{align}
  M_s &= A_1 s^2+B_1 s+\gamma_1, \label{intseq_a}\\
  L_s &= A_2 s^2+B_2 s+\gamma_2. \label{intseq_b}
  \end{align}
  \end{subequations}
  For $j>1$, the sequences $M_s$ and $L_s$ will take integer values for all $s$ if and only if $\gamma_i$ is an arbitrary integer and $A_i, B_i$ are both simultaneously either integers or half-integers for $i=1$ and $i=2$. The proof is elementary and we omit it. We shall henceforth assume that the above restriction ($j>1$) is verified and shall thus determine the FR conditions for generic chains with $N>3$. For very small chains, there are additional special possibilities that we shall not spell out here. With this understanding, we can cast the $A_i$ and $B_i$ as 
  \begin{subequations} \label{AB}
  \begin{align}
   A_1 = \frac{\alpha_1}{2}, \quad B_1=\frac{\beta_1}{2}, \\
   A_2 = \frac{\alpha_2}{2}, \quad B_2=\frac{\beta_2}{2}, 
  \end{align}
  \end{subequations}
  where $\alpha_1$ and $\beta_1$ are integers with the same parity and $\alpha_2$ and $\beta_2$ are also integers with the same parity. With \eqref{int_sequences} and \eqref{AB}, condition \eqref{arguments} becomes
  \begin{subequations} \label{arguments2}
  \begin{align}
  0=& (T-\pi\alpha_1)s^2+(2aT-\pi\beta_1)s\\&+(a^2T-\frac{\pi}{2}+2\theta+\phi-2\pi\gamma_1), \notag\\
  0=& (T-\pi\alpha_2)s^2+(2cT-\pi\beta_2)s\\&+(c^2T+\frac{\pi}{2}-2\theta+\phi-2\pi\gamma_2). \notag
  \end{align}
  \end{subequations}       
  Since these relations must hold for $s=0,1,\dots,j$, again with $j>1$, each coefficient must vanish. First, the coefficients of $s^2$ yield
  \begin{align} \label{time_alpha}
   T=\pi\alpha_1 = \pi\alpha_2,
  \end{align}
  which says that $\alpha_1=\alpha_2$ is a positive integer, that $\alpha_1, \beta_1$ and $\beta_2$ must share the same parity and that the time for FR to occur is a multiple of $\pi$. Second, in view of \eqref{time_alpha}, the coefficients of $s$ give 
  \begin{align} \label{paramAC}
  a = \frac{\beta_1}{2\alpha_1}, \quad c = \frac{\beta_2}{2\alpha_1}.
  \end{align}
  The parameters $a$ and $c$ are thus rational numbers. Recall that $a>-\frac{1}{2}$ and $|a|<c<|a+1|$ implying that
  \begin{align} \label{beta_cond}
  \beta_1 > -\alpha_1, \quad |\beta_1|< \beta_2 < |\beta_1 + 2\alpha_1|. 
  \end{align}
  Third, equating the constant terms provides
  \begin{subequations}
  \begin{align}
  a^2 T &= \frac{\pi}{2}-\phi-2\theta+2\pi\gamma_1, \\
  c^2 T &=-\frac{\pi}{2}-\phi+2\theta+2\pi\gamma_2. 
  \end{align}
  \end{subequations}
  It is helpful to consider the sum and difference of these two equations:
  \begin{subequations}
  \begin{align}
   T(c^2-a^2)&= -\pi+4\theta+2\pi(\gamma_2-\gamma_1), \label{cond1} \\
   T(c^2+a^2)&= -2\phi+2\pi(\gamma_1+\gamma_2). \label{cond2}
  \end{align}
  \end{subequations}
  Using \eqref{time_alpha} and \eqref{paramAC}, equation \eqref{cond1} provides a condition on the parameters $\beta_1$,$\beta_2$ and $\theta$:
  \begin{align} \label{spectrum_cond2}
  \frac{(\beta_2-\beta_1)(\beta_1+\beta_2)}{4}= \left[2(\gamma_2-\gamma_1)-1+\frac{4\theta}{\pi}\right]\alpha_1,
  \end{align}
  where $\beta_1,\beta_2,\gamma_1,\gamma_2$ are integers and $\alpha_1$ is a positive integer. Note that since $\beta_1$ and $\beta_2$ share the same parity, the LHS of \eqref{spectrum_cond2} is an integer. This implies that $\theta$ must be a multiple of $\frac{\pi}{4\alpha_1}$ :
  \begin{align} \label{theta}
   \theta = \frac{\pi p}{4q}
  \end{align}
  where $\frac{p}{q}$ is an irreducible fraction and $q$ divides $\alpha_1$. Finally, upon substituting \eqref{time_alpha} and \eqref{paramAC} in \eqref{cond2}, one obtains that the global phase $\phi$ is a fraction of $\pi$. Hence, the Hamiltonian \eqref{Hamiltonian} with couplings and magnetic fields given by \eqref{Couplings} will admit fractional revival if the parameters $a,c$ are of the form \eqref{paramAC} with $\beta_1,\beta_2,\alpha_1$ solutions of \eqref{spectrum_cond2} respecting \eqref{beta_cond}. An example of solution is : $\beta_1 = 14$, $\beta_2=16$, $\alpha_1=6$, $\theta=\frac{3\pi}{8}$ and $\gamma_2-\gamma_1 = 1$ which gives $a=\frac{7}{6}$ and $c=\frac{8}{6}$.

  Mirror-symmetry which is realized in the models we have been discussing so far is a necessary condition for PST. It is hence of interest to enquire if PST can also be observed in spin chains that have been found to exhibit FR. Once one has \eqref{15} with $\xi$ and $\eta$ given by \eqref{18}, it can be shown \cite{Fractional_quantum_spin_chains} that one has also
  \begin{align} \label{32}
   e^{-iMTJ}\ket{0} = e^{i\phi}\Big[ &\cos M\left(\frac{\pi}{2}-2\theta\right)\ket{0}\\ \notag &+ i\sin M\left(\frac{\pi}{2}-2\theta\right)\ket{N} \Big]
  \end{align}
  with $M$ an integer. This simply follows from formulas for $e^{-iTJ}$ that will be given for $N$ odd and $N$ even at the beginning of section 6. Perfect state transfer will then occur at time $MT$ if 
  \begin{align} \label{33}
   M\left(\frac{\pi}{2}-2\theta\right) = M_1 \left(\frac{\pi}{2}\right)
  \end{align}
  with $M_1$ an arbitrary odd number. Recall that \eqref{spectrum_cond2} requires \eqref{theta}. For fixed $p$ and $q$, given a set of parameters solving \eqref{spectrum_cond2}, we see that the corresponding model will possess PST in addition if
  \begin{align} \label{34}
   \frac{M_1}{M} = \frac{q-p}{q}.
  \end{align}
  This requires\footnote{The case $p=0$ is also possible with $M=M_1$ but in this instance there will just be replications of PST at times $M_1T$.} that $p$ and $q$ have opposite parities and $M=q$. This is ensured by the irreducibility of $\frac{p}{q}$ for $q$ even, but not for $q$ odd. Thus, spin chains with FR at time $T$ will also show PST at time $qT$ if $q$ is even but when $q$ is odd, PST will only happen at time $qT$ if $p$ is even. 

  It turns out that the solutions with PST can be identified with the models presented by Albanese and Lawi in an unpublished paper \cite{Albanese_Lawi_classification_2005} although the authors have assumed not quite correctly that the underlying polynomials are special cases of the Racah polynomials. 

  A particular feature of the para-Racah model is the bump in the couplings $J_n$ in the middle of the of chain. Here is a plot showing the strengths of the couplings along the chain :
  \vskip 5pt
  \noindent\includegraphics[height=5.0cm]{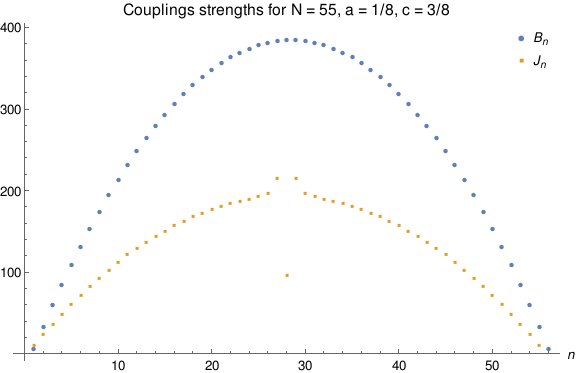}

\section{4. Spectral Surgery and the case N even}
  A procedure called spectral surgery and described in \cite{2012_Vinet&Zhedanov_PhysRevA_85_012323} allows to modify the spectrum of a Jacobi matrix while preserving its mirror-symmetry. It is hence possible to relate a chain with $N$ sites to one with $N-1$ sites by removing the last eigenvalue. That the modified chain will still enact fractional revival is obvious since the conditions \eqref{19} will remain satisfied for $s=0,1,\dots,N-1$.
  
  This technique hence permit to define our models for even $N$. The new couplings $\hat{B}_n,\hat{J}_n$ are related to \eqref{Couplings} by
  \begin{align}
  \hat{B}_{n}=B_{n+1}+A_{n+1}-A_n, \quad \hat{J}_{n}=J_n\left[\frac{A_n}{A_{n-1}}\right]^\frac{1}{2},
  \end{align} 
  where $A_n=\frac{P_{n+1}(-x_N;N;a,c,\frac{1}{2})}{P_n(-x_N;N;a,c,\frac{1}{2})}$ is given by
  \begin{align}
   A_n = \frac{(N-n)(n-j+a-c)(N-n-1+a+c)}{2(2n-N)}.
  \end{align}
  Upon changing $N\rightarrow N+1$, a direct computation yields the recurrence coefficients for $N=2j$:
  \begin{subequations} \label{CouplingsEven}
  \begin{align} 
   \hat{B}_n=& \frac{1}{2}(a^2+c^2+n-n^2)+\frac{1}{4}(2n+a+c)(N-1)\notag\\* &+\frac{(n+1)(n+a+c)(1+2a-2c)}{4(1+2n-N)}\\*\notag &+\frac{n(n-1+a+c)(1+2a-2c)}{4(1-2n+N)}
  \end{align}
  and
  \begin{align}
    \hat{J}_n=&\Big[\frac{n(N+1-n)(n-1+a+c)(N-n+a+c)}{4(N-2n+1)^2}\notag\\* &\times (n-j+a-c)(n-j+c-a-1)\Big]^\frac{1}{2}.
  \end{align}
  \end{subequations}
  It can be directly checked that the persymmetry condition \eqref{persymmetry} still holds while the spectrum of $J$ is now 
  \begin{align} \label{bi-lattice2}
  \begin{aligned}
    x_{2s}=&(s+a)^2, \quad s=0,\dots,j, \\
    x_{2s+1}=&(s+c)^2, \quad s=0,\dots,j-1.
  \end{aligned}
  \end{align}    
  Fractional revival will occur for $a$ and $c$ again given by \eqref{paramAC} and \eqref{spectrum_cond2}. Here, in view of \eqref{bi-lattice2} and of the remarks after equations \eqref{int_sequences}, we need to assume that $j>2$. Our analysis thus provide a full classification of the FR conditions for chains with $N>4$. The plot showing the strengths of the couplings along the chains exhibit a different behavior for $N$ even; here, both the couplings $J_n$ and the magnetic fields $B_n$ have a bump in the middle of the chains : 
  \vskip 8pt
  \noindent \includegraphics[height=5.0cm]{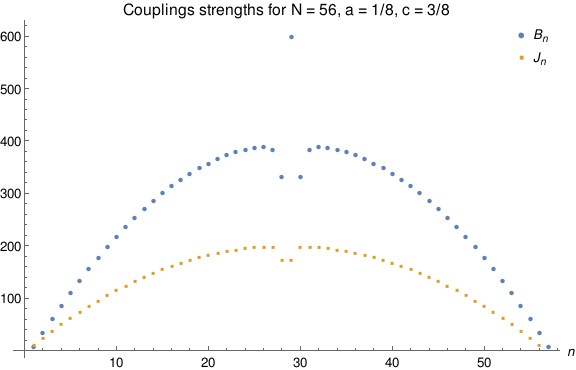} 

\section{5. Special case $c=a+\frac{1}{2}$}
 The models presented here encompass simpler known models. Note that upon setting $c=a+\frac{1}{2}$, the bi-lattice \eqref{bi-lattice} reduces to a single quadratic lattice of the form
  \begin{align}
  x_s = \left(\frac{s}{2}+a\right)^2.
  \end{align}
  In this case, the recurrence coefficients for odd $N$ \eqref{Couplings} and even $N$ \eqref{CouplingsEven} need not be distinguished and are given by
  \begin{align*}
   B_n &= N+4N(a+n)+a^2-\frac{n^2}{2},\\ \notag
   J_n &= \left[\frac{n(n+2a-\frac{1}{2})(N-n+2a+\frac{1}{2})(N+1-n)}{16}\right]^\frac{1}{2}.
  \end{align*}
  This correspond to the analytic spin chain models connected to the dual-Hahn polynomials that were identified by Albanese \emph{et al.} in \cite{2004_Albanese&Christandl&Datta&Ekert_PhysRevLett_93_230502}. Note that our analysis shows that these simpler models also exhibit FR for some values of the parameters. In this case, the bump in the middle couplings disappear and the patterns are those of smooth parabolas :
  \vskip 8pt
  \noindent \includegraphics[height=5.0cm]{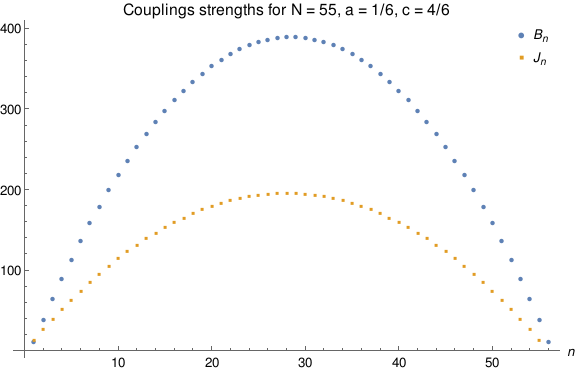}

\section{6. Isospectral deformations}
We shall now indicate how a second parameter in addition to $\theta$ can be introduced in the amplitudes of the two revived clones by means of an isospectral deformation.

Let us first point out that the conditions \eqref{19} do not only imply \eqref{15} with $\xi$ and $\eta$ as in \eqref{18} but more generally that
\begin{equation*}
 e^{-iTJ} = e^{i\phi} 
  \begin{pmatrix}
  \sin 2\theta &&&&& i\cos 2\theta \\
  &\ddots &&&\iddots & \\
  &&\sin 2\theta& i\cos2\theta&&\\
  &&i\cos 2\theta& \sin2\theta &&\\
  &\iddots &&&\ddots&\\
  i\cos 2\theta &&&&&\sin 2\theta
  \end{pmatrix}
\end{equation*}
for $N$ odd and
\begin{align*}
 &e^{-iTJ} = 
 \\
   \notag &e^{i\phi}\begin{pmatrix}
  \sin 2\theta  &&&&&& i\cos 2\theta\\
  & \ddots &&&&\iddots&\\
  &&\sin2\theta&0&i\cos2\theta &&\\
  && 0 & e^{i(\frac{\pi}{2}-2\theta)} &0 &&\\
  &&i\cos2\theta & 0 & \sin2\theta &&\\
  &\iddots &&&&\ddots&\\
  i\cos 2\theta &&&&&&\sin2\theta
  \end{pmatrix}   
\end{align*}
for $N$ even. This is not difficult to show and has been derived in \cite{Fractional_quantum_spin_chains}.

Now introduce the symetric matrix 
  \begin{align}
  V=
  \begin{pmatrix}
  \sin \sigma &&&&&\cos \sigma \\
  &\ddots &&&\iddots & \\
  &&\sin \sigma& \cos\sigma&&\\
  &&\cos \sigma& -\sin\sigma &&\\
  &\iddots &&&\ddots&\\
  \cos \sigma &&&&&-\sin \sigma
  \end{pmatrix},
  \end{align}
  for $N$ odd and
  \begin{align}
  V=
  \begin{pmatrix}
  \sin \sigma &&&&&&\cos\sigma\\
  & \ddots &&&&\iddots&\\
  &&\sin\sigma&0&\cos\sigma &&\\
  && 0 & 1 &0 &&\\
  &&\cos \sigma & 0 & -\sin\sigma &&\\
  &\iddots &&&&\ddots&\\
  \cos \sigma &&&&&&-\sin\sigma
  \end{pmatrix},
  \end{align}
  for $N$ even. It is easy to verify that $V^2=1$. Now let $J$ be a persymmetric Jacobi matrix and define $\tilde{J}$ via the conjugation
  \begin{align} \label{Conjugation}
  \tilde{J}=VJV.
  \end{align}
  It is clear that this provides an isospectral deformation of $J$. $\tilde{J}$ will no longer be persymmetric but will satisfy
  \begin{align}
   \tilde{J} = Q\tilde{J}Q
  \end{align}
  with $Q = VRV$. A direct computation shows that $e^{-iT\tilde{J}}$ acting on $\ket{0}$ gives
  \begin{align}
  e^{-iT\tilde{J}}\ket{0} = Ve^{-iTJ}V\ket{0} = \xi \ket{0} + \eta \ket{N} 
  \end{align}
  for both $N$ odd and $N$ even with 
  \begin{subequations} \label{general_param}
  \begin{align}
  &\xi = e^{i\phi}(\sin2\theta +2i\cos2\theta\cos\sigma\sin\sigma) \\
  &\eta = ie^{i\phi}\cos2\theta(\cos^2\sigma - \sin^2\sigma). 
  \end{align}
  \end{subequations}
  An additional angle $\sigma$ has thus been introduced in the parameterization of $\xi$ and $\eta$ and it is easily verified that the normalization condition $|\xi|^2+|\eta|^2=1$ is still satisfied. It is observed that $\tilde{J}$ is also tridiagonal, it thus provides the couplings and magnetic fields of an XX spin chain that exhibits again fractional revival at two sites. Remarkably, most of the coupling constants and magnetic fields of $J$ remains unchanged. Carrying out the transformation \eqref{Conjugation}, it is seen that the only entries of $\tilde{J}$ that differ from those of $J$ are
  \begin{subequations} \label{48}
  \label{Perturbations}
  \begin{align}
  \begin{aligned}
  \widetilde{J}_{\frac{N+1}{2}}&=J_{\frac{N+1}{2}}\,\cos 2\sigma,
  \\
  \widetilde{B}_{\frac{N\mp 1}{2}}&= B_{\frac{N-1}{2}}\pm J_{\frac{N+1}{2}}\sin 2\sigma,
  \end{aligned}
  \end{align}
  for $N$ odd and
  \begin{align}
  \begin{aligned}
  \widetilde{J}_{\frac{N}{2}}&= J_{\frac{N}{2}}(\cos \sigma-\sin \sigma),
  \\
  \widetilde{J}_{\frac{N}{2}+1}&=J_{\frac{N}{2}}(\cos \sigma+\sin \sigma),
  \end{aligned}
  \end{align}
  \end{subequations}
  for $N$ even. Remembering the expressions \eqref{Couplings} and \eqref{CouplingsEven} for the entries of $J$, it is seen that the elements of $\tilde{J}$ ($\tilde{J}_n$ and $\tilde{B}_n$) correspond to the recurrence coefficients of the para-Racah polynomials $P_n(y^2;N;a,c,\alpha)$ \cite{Para_Racah_Polynomials} with $\sin(2\sigma)=1-2\alpha$ for $N$ odd and  $\sin\sigma = \frac{\sqrt{\alpha}-\sqrt{1-\alpha}}{\sqrt{2}}$ for $N$ even. We thus observe that the general para-Racah polynomials are associated with XX spin chains with generic fractional revival described by two parameters provided that $a$ and $c$ remain given by \eqref{paramAC} in terms of solutions of \eqref{spectrum_cond2}. 

\section{7. Conclusion}
  Let us summarize our findings and make a few additional remarks to conclude. We have introduced a novel analytic XX spin chain with fractional revival. It depends on 3 parameters $a$, $c$, $\alpha$ in addition to the number of sites $N+1$. Its nearest-neighbor couplings and local magnetic field strengths are provided by the recurrence coefficients of the recently identified para-Racah polynomials $P_n(x^2;N;a,c,\alpha)$ which are orthogonal on the quadratic bi-lattice \eqref{bi-lattice} characterized by $a$ and $c$. When $c=a+\frac{1}{2}$, these $P_n$ reduce to the dual-Hahn polynomials. There are constraints on the parameters for fractional revival to take place after time $T$. One must have according to \eqref{time_alpha} and \eqref{paramAC} : 
  \begin{align}
   a = \beta_1 \frac{\pi}{2T} \quad c = \beta_2 \frac{\pi}{2T}
  \end{align}
  where $\beta_1$ and $\beta_2$ are integers that solve the Diophantine equation \eqref{spectrum_cond2}. (There might be additional special solutions when $N\leq 4$). There are two angles $\theta$ and $\sigma$ that determine the FR amplitudes $\xi$ and $\eta$ in \eqref{15} as per the parameterization \eqref{general_param}. The first, $\theta$, is determined by \eqref{spectrum_cond2} and forced to take the restricted values specified by $\theta = \frac{\pi p}{4q}$ with $p, q$ co-primes and $q$ dividing $T/\pi$. The second, $\sigma$, is directly related to $\alpha$ in a way that depends on whether $N$ is odd or even (see the formulas in the paragraph after eqs.\eqref{48}). When $\sigma=0$, that is when $\alpha = \frac{1}{2}$, the recurrence coefficients $J_n$ and $B_n$ form a mirror-symmetric matrix. A necessary condition for perfect state transfer is then realized and it has been indicated that if FR happens at time $T$, PST will occur at time $qT$ when the parameters $a$ and $c$ lead to an angle of the form $\theta = \frac{\pi p}{4q}$ where $q$ and $p$ have different parities. Solutions of \eqref{spectrum_cond2} with $\theta=0$ provide models that exhibit PST but not FR. 

  The study presented here complements and extends the analysis \cite{Fractional_quantum_spin_chains} of the transport properties of the spin chains connected to the para-Krawtchouk polynomials \cite{2012_Vinet&Zhedanov_JPhysA_45_265304} which are orthogonal on the linear bi-lattice \eqref{1} instead of the quadratic bi-lattice \eqref{bi-lattice}. It is informative to compare the features of the latter model with those of the para-Racah system that we have examined so far in this paper. 

  To focus on the model with FR which is based on the para-Krawtchouk polynomials per se, we must set the angle $\sigma$ equal to zero. We are thus in a mirror-symmetric situation. For the para-Krawtchouk model, the FR angle $\theta$ is related in a simple way to the bi-lattice parameter $\delta$ in \eqref{1}, one has 
  \begin{align}
   \delta = 1 \pm \frac{4\theta}{\pi}.
  \end{align}
  There are no further conditions to have FR. However, in order for PST to manifest itself, we must have that
  \begin{align}
   \delta = \frac{M_1}{M}
  \end{align}
  where $M_1$ and $M$ are positive co-prime integers and $M_1$ is odd. 

  The para-Krawtchouk spin chains also possess another interesting property. When $\delta$ is an irrational number, even though state transfer is no longer perfect, it can be approached in finite time to any level of precision. It is indeed possible to show \cite{2012_Vinet&Zhedanov_PhysRevA_86_052319} using classical theorems of Diophantine approximations, that there exists a sequence of times $t_n$, $n=0,1,2,\dots$, such that
  \begin{align}
   \big|\bra{N}e^{-iJt_n}\ket{0}\big| \to 1
  \end{align}
  as $n\to\infty$. It follows that we have in these models, manifestations of what is called almost perfect state transfer (APST).

  These last comments suggest directions for future investigations. A first question is : under what circumstances would the mirror-symmetric spin chains associated to the para-Racah polynomials exhibit APST? Other questions relate to FR. We have been dealing in this paper with perfect fractional revival. Let $\big|\ket{\psi}\big|$ denote the norm of $\ket{\psi}$ and $\epsilon$ be a small positive real number. It would be of interest to determine the conditions for almost perfect fractional revival where
  \begin{align}
   \big|e^{-iTJ}\ket{0} - \left(\xi\ket{0}+\eta\ket{N}\right)\big| < \epsilon
  \end{align}
  or where, in other words, $e^{-iTJ}\ket{0}$ can be as close as desired to the state with two localized clones at $\ket{0}$ and $\ket{N}$. It would seem relevant to probe the para-Krawtchouk and para-Racah chains in this respect.  

\section*{Acknowledgments}
The authors wishes to acknowledge stimulating conversations with Leonardo Banchi and Vincent X. Genest. J.M.L. holds a scholarship from the Fonds de recherche du Qu\'ebec -- Nature et technologies (FRQNT). The research of L.V. is supported in part by NSERC. A. Z. wishes to thank the Centre de Recherches Math\'ematiques (CRM) for its hospitality.

\end{document}